\documentclass[
preprint,
 amsmath,amssymb,
 aps,
prb,
floatfix,
]{revtex4-2}

\usepackage{graphicx}
\usepackage{dcolumn}
\usepackage{booktabs, tabularx, ragged2e}
\usepackage{bm}
\usepackage[mathlines]{lineno}
\usepackage{commath}
\usepackage{physics}
\usepackage{color}
\usepackage{mathtools}

\begin{document}

\preprint{APS/123-QED}

\title{Strong unidirectional Rashba state induced by extended vacancy line defects in a $1T'$-WTe$_{2}$ monolayer}
 
\author{Moh. Adhib Ulil Absor}
\affiliation{Departement of Physics, Faculty of Mathematics and Natural Sciences, Universitas Gadjah Mada, Sekip Utara BLS 21 Yogyakarta 55186 Indonesia.}
\email{adib@ugm.ac.id}

\author{Iman Santoso}
\affiliation{Departement of Physics, Faculty of Mathematics and Natural Sciences, Universitas Gadjah Mada, Sekip Utara BLS 21 Yogyakarta 55186 Indonesia.}

\author{Harsojo}
\affiliation{Departement of Physics, Faculty of Mathematics and Natural Sciences, Universitas Gadjah Mada, Sekip Utara BLS 21 Yogyakarta 55186 Indonesia.}


\date{\today}

\begin{abstract}
The correlation between spin-orbit coupling and low crystal symmetry in the $1T'$ phase of the tungsten ditellurides (WTe$_{2}$) monolayer (ML) plays a significant role in its electronic and topological properties. However, the centrosymmetric nature of the crystal maintains Kramer's spin degeneracy in its electronic states, which limits its functionality in spintronics. In this paper, through a systematic study using first-principles calculations, we show that significant spin splitting can be induced in the $1T'$-WTe$_{2}$ ML by introducing one dimensional (1D) vacancy line defect (VLD). We examine six configurations of the 1D VLD, which consist of three VLDs extended in the armchair direction including a Te$_{1}$ armchair-VLD ($ACV_{\texttt{Te}_{1}}$), Te$_{2}$ armchair-VLD ($ACV_{\texttt{Te}_{2}}$), and W armchair-VLD ($ACV_{\texttt{W}}$); and three VLDs elongated along the zigzag direction comprising a Te$_{1}$ zigzag-VLD ($ZZV_{\texttt{Te}_{1}}$), Te$_{2}$ zigzag-VLD ($ZZV_{\texttt{Te}_{2}}$), and W zigzag-VLD ($ZZV_{\texttt{W}}$), where Te$_{1}$ and Te$_{2}$ are two nonequivalent Te atoms located at the lower and higher sites in the top layer, respectively. We find that both the $ACV_{\texttt{Te}_{1}}$ and $ACV_{\texttt{W}}$ systems have the lowest formation energy. Concerning these two most stable VLD systems, we identify large spin splitting in the defect states near the Fermi level driven by a strong coupling of the in-plane $p-d$ orbitals, displaying highly unidirectional Rashba states with perfectly collinear spin configurations in the momentum space. This unique spin configuration gives rise to a specific spin mode that protects the spin from decoherence and leads to an exceptionally long spin lifetime. Furthermore, the observed unidirectional Rashba states are enforced by the inversion symmetry breaking and the 1D nature of the VLD, as clarified by the $\vec{k}\cdot\vec{p}$ model derived from the symmetry analysis. Our findings pave a possible way to induce significant spin splitting in the $1T'$-WTe$_{2}$ ML, which could be highly important for designing highly efficient spintronic devices.
\end{abstract}

\pacs{Valid PACS appear here}
\maketitle

\section{INTRODUCTION}

Two-dimensional (2D) layered materials arranged through van der Waals forces have attracted substantial interest for recent years due to their fascinating characteristics and immense promise across (opto)electronics \cite{Cui2015, Cheng2014, Wang2012}, spintronics \cite{Avsar}, as well as energy conversion \cite{Kononov2020, MacNeill2017}. Among these 2D layered materials, transition metal dichalcogenides (TMDCs) hold a notable portion \cite{Manzeli2017}, featuring a range of $MX_{2}$ compositions where $M$ stands for a transition metal element and $X$ represents a chalcogen element. Weak van der Waals interactions between successive $MX_{2}$ layers permit the separation of TMDCs into single layers through the method of mechanical exfoliation \cite{Pnas}. The common bulk phases of the TMDCs include $2H$, $1T$, and $3R$ phases involving stacking layers with hexagonal, octahedral, and rhombohedral symmetries, respectively \cite{Voiry, Patil, Manzeli2017}. While $2H$ and $3R$ phases exhibit trigonal prismatic coordination of metal atoms, the $1T$ phase has trigonal anti-prismatic or octahedral coordination. Moreover, the metastable $1T$ phase tends to spontaneously distort its lattice by dimerizing transition metal atoms along one direction, leading to reduced lattice symmetry. This structure features one-dimensional dimers of metal atoms flanked by zigzag chains of chalcogen atoms, exhibiting an inversion-symmetric $1T'$ and an inversion symmetry-broken $T_{d}$ stacking orders \cite{Qian2014, YTao, Singh}. In the monolayer (ML) limit, even though the stable $1H$ and $1T$ phases are present, the $1T'$ and $T_{d}$ phases also emerge, depending on the specific pairing of transition metal $M$ and chalcogen $X$ elements \cite{Qian2014, YTao, Singh}. The diverse combinations of chemical compositions and structural phases in TMDCs ML lead to a wide spectrum of electronic properties, which encompasses variations in band structure character (metallic or insulating) \cite{Manzeli2017, Zhao2020} and the emergence of correlated and topological phases \cite{Qian2014, Zhao2020, Xu2018}. This underscores the suitability of TMDCs ML as an excellent platform for future-generation technologies.

An important characteristic discovered within TMDCs ML pertains to the robust influence of spin-orbit coupling (SOC), prominently observed in the $1H$ phase, as seen in molybdenum and tungsten dichalcogenides \cite{Zhu, Kosminder, Absor2016, Affandi, Guo, Bragan}. In this context, the absence of inversion symmetry in the crystal arrangement, coupled with the strong SOC exhibited by the 5$d$ orbitals of the transition metal atoms, gives rise to a significant spin splitting within the electronic band structures. This spin splitting is believed to underlie intriguing phenomena in the TMDCs ML, including the spin Hall effect \cite{Cazalilla, Qian2014}, selective spin-dependent rules for optical transitions \cite{Chu}, and the magneto-electric effect \cite{Gong}. Additionally, owing to the in-plane mirror symmetry present in the crystal structures of TMDCs ML, a distinctive out-of-plane spin polarization is retained within the spin-split bands surrounding the $K$ point within the first Brillouin zone (FBZ), resulting in a Zeeman-type spin splitting \cite{Zhu, Kosminder, Absor2016, Bragan}. This phenomenon is projected to display substantial spin coherence and prolonged spin relaxation for electrons \cite{Bragan}. Furthermore, recent reports highlight the feasibility of electrically manipulating spin splitting and spin polarization in the TMDCs ML, rendering them well-suited for applications in spintronics devices like spin-field effect transistors (SFET)\cite{Radisavljevic}. 

In comparison to the $1H$ phase of TMDCs ML, the influence of the SOC in the $1T'$ phases presents an equally captivating aspect. Particularly, the ML form of tungsten ditellurides (WTe$_{2}$) in the $1T'$ structure has garnered considerable scientific interest following its successful synthesis \cite{Zheng2016, Tang2017, Naylor_2017, Zhao, Zhao2020, Tan2021, Xu2018}. This ML embodies a complex interplay between the arrangement of bands linked to the SOC and the characteristics of W $d$-orbitals, which shed light on electronic correlations \cite{Zheng2016, Tang2017, Zhao2020, Xu2018, Xie, Shi, Garcia, Garcia_A, Vila}. Recent theoretical predictions \cite{Garcia, Garcia_A} and experimental observations \cite{Zhao} have unveiled the emergence of a canted quantum spin Hall effect within the WTe$_{2}$ ML, driven by a canted unidirectional SOC. This discovery adds a new dimension to the potential utilization of topological materials in spintronic applications \cite{Garcia_A, Vila}. In addition, the interplay of the SOC, low crystal symmetry, and high electron mobilities give rise to other phenomena of interest such as large spin-orbit torques \cite{MacNeill2017, Li2018} that could be used in spintronic devices. However, it's worth noting that in the $1T'$ phase of the WTe$_{2}$ ML, the electronic band structures maintain spin degeneracy due to the crystal's centrosymmetric nature, which imposes limitations on its functionality in spintronics. Given the WTe$_{2}$ ML's remarkable transport properties \cite{Zheng2016, Tang2017, Zhao, Tan2021, Xu2018, Xie, Garcia, Garcia_A, Vila}, the key to enabling its use in spintronic devices likely lies in overcoming this spin degeneracy. Thus, identifying a viable approach to induce substantial spin splitting within the $1T'$ phase of WTe$_{2}$ ML becomes a highly desirable pursuit.

In this paper, by using density-functional theory (DFT) calculations, we report that significant spin splitting can be induced in the $1T'$-WTe$_{2}$ ML by introducing one dimensional (1D) extended vacancy line defect (VLD). Previous research has already highlighted the significant impact of surface imperfections on the electronic and topological properties of the $1T'$-WTe$_{2}$ ML such as point defects \cite{Ozdemir2022, Muechler, Song2018, Chen2022} and 1D edge step surfaces \cite{Peng2017, Lau}. Since the emergence of the vacancy defects in the $1T'$-WTe$_{2}$ ML has been experimentally reported through scanning tunneling microscopy study \cite{Peng2017}, this suggests that realization of the VLD in the $1T'$-WTe$_{2}$ ML is highly plausible. In our study, we consider six configurations of the 1D VLD, which consist of three VLDs extended in the armchair direction and three VLDs elongated along the zigzag direction. The armchair-VLDs include a Te$_{1}$ armchair-VLD ($ACV_{\texttt{Te}_{1}}$), Te$_{2}$ armchair-VLD ($ACV_{\texttt{Te}_{2}}$), and W armchair-VLD ($ACV_{\texttt{W}}$), while the zigzag-VLDs comprise a Te$_{1}$ zigzag-VLD ($ZZV_{\texttt{Te}_{1}}$), Te$_{2}$ zigzag-VLD ($ZZV_{\texttt{Te}_{2}}$), and W zigzag-VLD ($ZZV_{\texttt{W}}$), where Te$_{1}$ and Te$_{2}$ are two nonequivalent Te atoms located at the lower and higher sites in the top layer, respectively. Our findings reveal that both the $ACV_{\texttt{Te}_{1}}$ and $ACV_{\texttt{W}}$ have the lowest formation energy. By focusing on these two most stable VLDs, we observe a significant spin splitting in the defect states near the Fermi level imposed by a strong hybridization of the in-plane $p-d$ orbitals, exhibiting a highly unidirectional Rashba states with perfectly collinear spin configurations in the momentum space. These particular spin configurations result in a specific spin mode that protects the spin from decoherence and results in an extremely long spin lifetime \cite{Dyakonov, Schliemann, Bernevig, SchliemannJ, Altmann}. Furthermore, the observed unidirectional Rashba states are enforced by the inversion symmetry breaking and the 1D nature of the defect, as confirmed by the $\vec{k}\cdot\vec{p}$ model derived from the symmetry analysis. Finally, a possible application of the present system for spintronics will be discussed.

\section{Computational Details}

All DFT calculations were performed based on norm-conserving pseudo-potentials and optimized pseudo-atomic localized basis functions implemented in the OpenMX code \cite{Ozaki, Ozakikino, Ozakikinoa}. The exchange-correlation function was treated within generalized gradient approximation by Perdew, Burke, and Ernzerhof (GGA-PBE) \cite{gga_pbe, Kohn}. The basis functions were expanded by a linear combination of multiple pseudo atomic orbitals (PAOs) generated using a confinement scheme \cite{Ozaki, Ozakikino}, where three $s$-, three $p$-, two $d$-character numerical PAOs were used. The accuracy of the basis functions, as well as pseudo-potentials we used, were carefully bench-marked by the delta gauge method \cite{Lejaeghere}.

\begin{figure}
	\centering
		\includegraphics[width=1.0\textwidth]{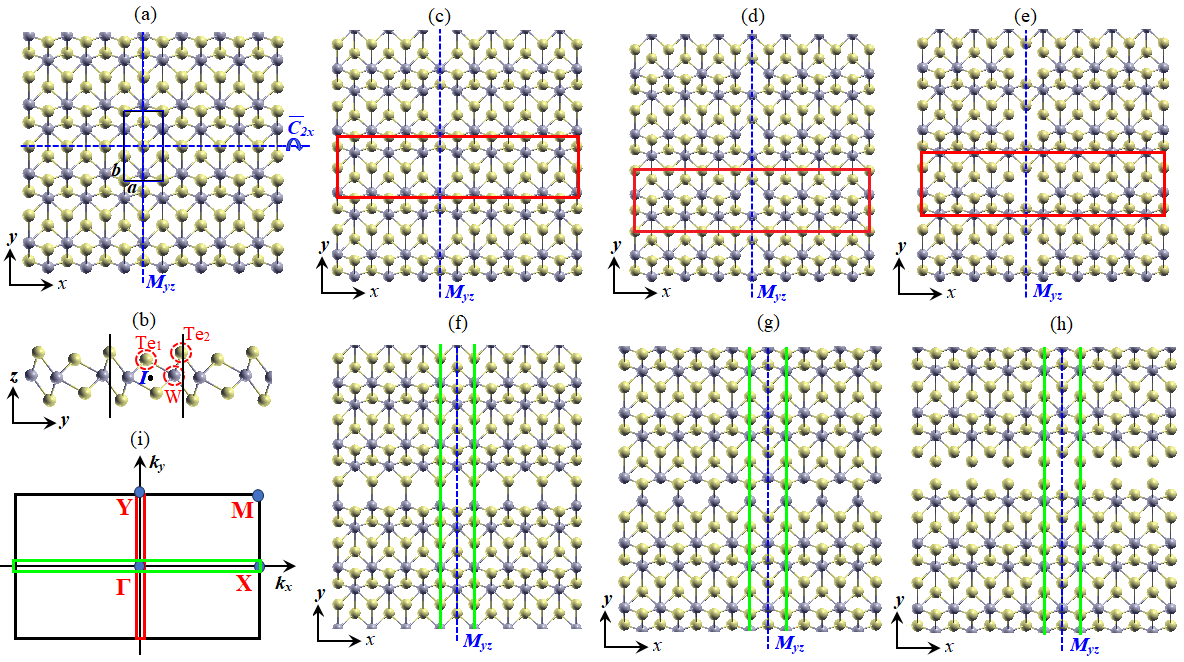}
	\caption{(a)-(b) Top and side views of the pristine $1T'$-WTe$_{2}$ ML is shown, respectively. The black line represents the primitive rectangular unit cell with $a$ and $b$ lattice parameters along the $x$ and $y$ direction, respectively. Here, two nonequivalent Te sites (Te$_{1}$ and Te$_{2}$) distinguished by the lower and higher sites in the top layer of the $1T'$-WTe$_{2}$ ML, respectively, are indicated. Top and side views for the 1D VLD supercell with (c) a Te$_{1}$ armchair-VLD ($ACV_{\texttt{Te}_{1}}$), (d) a Te$_{2}$ armchair-VLD ($ACV_{\texttt{Te}_{2}}$), (e) a W armchair-VLD ($ACV_{\texttt{W}}$), (f) a Te$_{1}$ zigzag-VLD ($ZZV_{\texttt{Te}_{1}}$), (g) a Te$_{2}$ zigzag-VLD ($ZZV_{\texttt{Te}_{2}}$), and (h) a W zigzag-VLD ($ZZV_{\texttt{W}}$) configurations are shown. The red and green rectangles indicate the 1D supercell with the VLD extended along the armchair and zigzag directions, respectively. (i) The FBZ of the primitive rectangular unit cell and the corresponding 1D armchair-VLD (red lines) and zigzag-VLD (green lines) supercells are shown. The symmetry operations including inversion symmetry ($I$), a mirror symmetry ($M_{yz}$) perpendicular to the $x$ axis, and a two-fold screw rotation symmetry ($\bar{C}_{2x}$) around the $x$ axis are indicated. }
	\label{figure:Figure1}
\end{figure}

To simulate the VLD, we constructed a supercell of the pristine $1T'$-WTe$_{2}$ ML by using the optimized primitive rectangular unit cell [Figs. 1(a)-(b)]. Here, we used the axis system where the $1T'$-WTe$_{2}$ ML is chosen to sit on the $x-y$ plan, where the $x$ ($y$) axis was oriented parallel to the zigzag (armchair) direction. We explore six distinct configurations of the 1D VLD, comprising three VLDs extended in the armchair direction and three in the zigzag direction. Specifically, for the armchair-oriented 1D VLD, we examine three configurations, namely a Te$_{1}$ armchair-VLD ($ACV_{\texttt{Te}_{1}}$), Te$_{2}$ armchair-VLD ($ACV_{\texttt{Te}_{2}}$), and W armchair-VLD ($ACV_{\texttt{W}}$), as depicted in Figs. 1(c)-1(e), respectively. Conversely, in the case of the 1D VLD elongated along the zigzag direction, we analyze three configurations, including a Te$_{1}$ zigzag-VLD ($ZZV_{\texttt{Te}_{1}}$), Te$_{2}$ zigzag-VLD ($ZZV_{\texttt{Te}_{2}}$), and W zigzag-VLD ($ZZV_{\texttt{W}}$) as shown in Figs. 1(f)-1(h), respectively. Here, Te$_{1}$ and Te$_{2}$ are the Te atoms which are distinguished by their position in the lower and higher sites in the top layer of the $1T'$-WTe$_{2}$ ML, as shown in Fig. 1(b). The first Brillouin zone (FBZ) for the primitive unit cell and the corresponding supercell of the VLD systems is shown in Fig. 1(i). To represent these VLDs in our simulations, we enlarged the supercell size by a factor of 7 in the zigzag ($x$) and armchair ($y$) direction for the armchair-VLD and zigzag-VLD systems, respectively, to prevent interactions between periodic images of the line defects [see Figs. 1(c)-1(h)]. We used a periodic slab with a sufficiently large vacuum layer (20 \AA) to avoid interaction between adjacent layers. The $2\times8\times1$ $k$-point and $12\times2\times1$ $k$-point meshes were used for the armchair-VLD and zigzag-VLD systems, respectively. The geometries were fully relaxed until the force acting on each atom was less than 1 meV/\AA. 

To confirm the energetic stability of the VLD, we compute the formation energy $E^{f}$ through the following formula \cite {Freysoldt, Absor2020}:
\begin{equation}
\label{1}
E^{f}=E_{\texttt{VLD}}-E_{\texttt{Pristine}}+\sum_{i}n_{i}\mu_{i}.
\end{equation}
In Eq. (1), $E_{\texttt{VLD}}$ is the total energy of the VLD system, $E_{\texttt{Pristine}}$ is the total energy of the pristine $1T'$ WTe$_{2}$ ML supercell, $n_{i}$ is the number of atom being removed from the pristine system, and $\mu_{i}$ is the chemical potential of the removed atoms corresponding to the chemical environment surrounding the system. Here, $\mu_{i}$ obtains the following requirements:
\begin{equation}
\label{2a}
E_{\texttt{WTe}_{2}}-2E_{\texttt{Te}}\leq \mu_{\texttt{W}}\leq E_{\texttt{W}},
\end{equation}
\begin{equation}
\label{2b}
\frac{1}{2}(E_{\texttt{WTe}_{2}}-E_{\texttt{W}})\leq \mu_{\texttt{Te}}\leq E_{\texttt{Te}}.
\end{equation}
Under Te-rich condition, $\mu_{\texttt{Te}}$ is the energy of the Te atom in the bulk phase which corresponds to the lower limit on W, $\mu_{\texttt{W}}=E_{\texttt{WTe}_{2}}-2E_{\texttt{Te}}$, where $E_{\texttt{WTe}_{2}}$ is the total energy of the $1T'$-WTe$_{2}$ ML in the primitive unit cell. On the other hand, in the case of the W-rich condition, $\mu_{\texttt{W}}$ is associated with the energy of the W atom in the bulk phase corresponding to the lower limit on Te, $\mu_{\texttt{Te}}=\frac{1}{2}(E_{\texttt{WTe}_{2}}-E_{\texttt{W}})$.

To evaluate electronic properties of the pristine and defective $1T'$-WTe$_{2}$ ML, we calculate band structures corresponding to the orbital-resolved projected bands along the specific $k$-path of the FBZ as depicted in Fig. 1(i). We then calculate spin-resolved projected bands by evaluating the spin vector component ($S_{x}$, $S_{y}$, $S_{z}$) in the $k$-space by using the spin density matrix after self-consistent field is achieved in the DFT calculations \cite{Kotaka}. The spin density matrix, denoted as $P_{\sigma \sigma^{'}}(\vec{k},\mu)$, is computed using the spinor Bloch wave function, $\Psi^{\sigma}_{\mu}(\vec{r},\vec{k})$, through the following equation, 
\begin{equation}
\begin{aligned}
\label{2c}
P_{\sigma \sigma^{'}}(\vec{k},\mu)=\int \Psi^{\sigma}_{\mu}(\vec{r},\vec{k})\Psi^{\sigma^{'}}_{\mu}(\vec{r},\vec{k}) d\vec{r}\\
                                  = \sum_{n}\sum_{i,j}[c^{*}_{\sigma\mu i}c_{\sigma^{'}\mu j}S_{i,j}]e^{\vec{R}_{n}\cdot\vec{k}},\\
\end{aligned}
\end{equation}
where $\Psi^{\sigma}_{\mu}(\vec{r},\vec{k})$ is obtained after self-consistent is achieved in the DFT calculation. In Eq. (\ref{1}), $S_{ij}$ is the overlap integral of the $i$-th and $j$-th localized orbitals, $c_{\sigma\mu i(j)}$ is expansion coefficient, $\sigma$ ($\sigma^{'}$) is the spin index ($\uparrow$ or $\downarrow$), $\mu$ is the band index, and $\vec{R}_{n}$ is the $n$-th lattice vector. 

\section{RESULT AND DISCUSSION}

Before examining the defective systems, we first provide a brief overview of the structural symmetry and electronic properties of the pristine system. The ML structure of WTe$_{2}$ adopts a stable $1T'$ structure, in which the W atoms are organized in octahedral coordination with the Te atoms. This arrangement leads to a slightly buckled zigzag chain of W atoms due to metallic bonding, resulting in a distortion of the Te octahedron around each Te atom [Figs. 1(a)-(b)]. The crystal structure of the $1T'$-WTe$_{2}$ ML is centrosymmetric and its symmetry belongs to the $C_{2h}$ point group, characterized by several symmetry operations including identity ($E$), inversion symmetry ($I$), a mirror symmetry ($M_{yz}$) perpendicular to the $x$ axis, and a two-fold screw rotation symmetry ($\bar{C}_{2x}$) around the $x$ axis, as displayed in Figs. 1(a)-(b). The optimized lattice parameters $a$ and $b$ of the $1T'$-WTe$_{2}$ ML are found to be 3.51 \AA\ and 6.23 \AA, respectively. Nevertheless, these results are consistent with previous theoretical reports \cite{Zheng2016, Torun, RamanYang, Ozdemir2022} and experimental data \cite{Zheng2016, Tang2017, Naylor_2017}. Notably, the substantial difference between the $a$ and $b$ parameters signifies pronounced anisotropy in the crystal geometry, implying that this material exhibits highly anisotropic electronic, mechanical, and optical properties when subjected to uniaxial strain along the $x$- and $y$-directions \cite{Torun, RamanYang}.

\begin{figure*}
	\centering
		\includegraphics[width=1.0\textwidth]{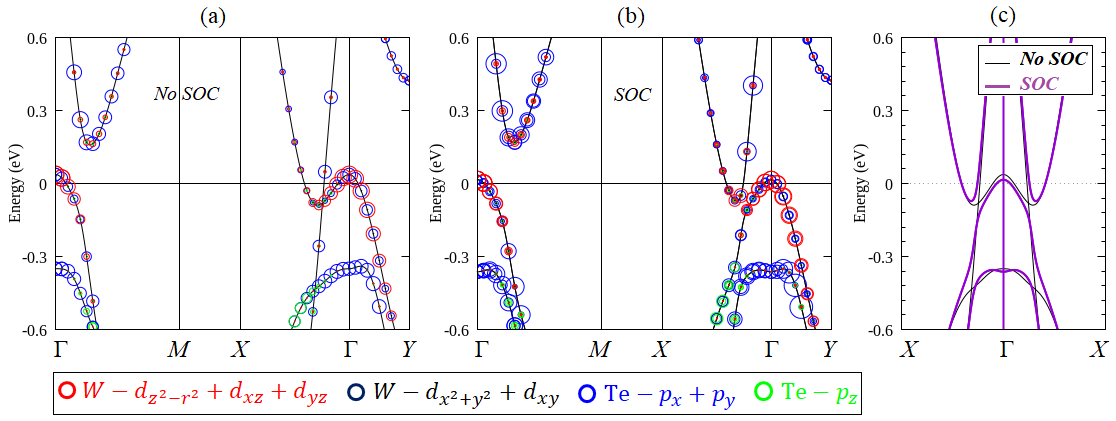}
	\caption{(a)-(b) Orbital-resolved of the electronic band structure of the pristine ($1\times 1$) $1T'$-WTe$_{2}$ ML calculated without and with the spin-orbit coupling (SOC), respectively. The orbital atomic components are indicated by the color circles, where the radii of the circles reflect the magnitude of the spectral weight of the particular orbitals to the bands. (c) Band dispersion of the pristine ($1\times 1$) $1T'$-WTe$_{2}$ ML along the $X-\Gamma-X$ line calculated with (red lines) and without (black lines) the SOC is highlighted.}
	\label{figure:Figure2}
\end{figure*}

Figs. 2(a) and 2(b) show the orbital-resolved electronic band structure of the pristine (1x1) $1T'$-WTe$_{2}$ ML along the selected $\vec{k}$ path in the FBZ calculated with and without the SOC, respectively, while Fig. 2(c) highlight the band dispersion near the Fermi level along the $X-\Gamma-X$ line. In the absence of the SOC, one can see that the $1T'$-WTe$_{2}$ ML show a semi-metallic character of the electronic states, exhibiting Dirac crossing bands between the highest occupied state (HOS) and lowest unoccupied state (LUS) at the $k$ point along the $X-\Gamma$ line near the Fermi level [Figs. 2(a) and 2(c)]. These crossing bands are protected by $\bar{C}_{2x}$ screw rotation as well as $I$ inversion symmetries of the $C_{2h}$ point group. Due to the close electronegativity between the W and Te atoms, the W-$d_{z^{2}-r^{2}}+d_{xz}+d_{yz}$ states are strongly hybridized with the Te-$p_{x}+p_{y}$ states around the Fermi level [Fig. 2(b)], which is in contrast to the widely studied $1H$-$MX_{2}$ TMDCs ML, where the band gap occurs between the occupied transition metal $M-d$ and unoccupied chalcogen $X-p$ states \cite{Zhu, Kosminder, Absor2016, Affandi, Guo, Bragan}. With the inclusion of SOC, the bands further hybridize with each other due to the strong mixing between W-$d_{z^{2}-r^{2}}+d_{xz}+d_{yz}$ and Te-$p_{x}+p_{y}$ states, and the degeneracies at the Dirac crossing points are lifted [Figs. 2(b) and 2(c)]. Accordingly, opening a band gap in the bulk states occurs, thus the $1T'$-WTe$_{2}$ ML becomes a 2D topological insulator \cite{Zhao2020, Xu2018, Xie, Shi, Garcia, Garcia_A, Vila}. However, due to the presence of the inversion symmetry in the $1T'$-WTe$_{2}$ ML, all the bands remain degenerated in the entirely FBZ [see Figs. 2(b) and 2(c)]. 

We emphasized here that the absence of the spin-splitting bands plays a detrimental role in the spintronic properties of the $1T'$-WTe$_{2}$ ML. Therefore, we applied the 1D VLD to break the inversion symmetry of the $1T'$-WTe$_{2}$ ML, which is expected to induce the substantial spin-splitting bands. In fact, the defect inducing large band splitting has been previously reported on the $1H$ and $1T$ types of the TMDCs MLs \cite{Absor2020, Li2019, Absor2017, Li, Absor2018}. 

\begin{table}[ht!]
\caption{The calculated formation energy (measured in eV) of the Te-rich ($E^{f}_{\texttt{Te-rich}}$) and W-rich ($E^{f}_{\texttt{W-rich}}$) for the VLD systems corresponding to the W-Te$_{1}$ bond length ($d_{\textbf{W-Te}_{1}}$, in \AA) and the W-Te$_{2}$ bond length ($d_{\textbf{W-Te2}}$, in \AA) around the VLD site compared with that of the pristine systems. We also provide the calculated results of the single vacancy defect (SVD) such as Te$_{1}$ SVD ($V^{\texttt{SVD}}_{\texttt{Te}_{1}}$), Te$_{2}$ SVD ($V^{\texttt{SVD}}_{\texttt{Te}_{2}}$), and W SVD ($V^{\texttt{SVD}}_{\texttt{W}}$), and compare these results with those obtained from the previous reports for a comparison.} 
\centering 
\begin{tabular}{cc cc cc cc cc cc cc} 
\hline\hline 
Defective systems && ($E^{f}_{\texttt{Te-rich}}$; $E^{f}_{\texttt{We-rich}}$) (eV) && $d_{\textbf{W-Te}_{1}}$ (\AA) && $d_{\textbf{W-Te}_{2}}$ (\AA) &&  Reference \\ 
\hline 
\textbf{Pristine}                            &&                && 2.74     && 2.84           &&  This work \\
                                    &&                && 2.72     && 2.82           &&  Ref.\cite{Ozdemir2022} \\
\textbf{Vacancy line defect (VLD)}           &&                &&          &&                &&        \\
\textbf{Armchair-VLD}           &&                &&          &&                &&        \\
$ACV_{\texttt{Te}_{1}}$               && (2.37; 2.53)   && 2.73     && 2.85               &&  This work\\
$ACV_{\texttt{Te}_{2}}$               && (2.86; 3.06)   && 2.76     && 2.91           &&  This work\\ 
$ACV_{\texttt{W}}$                    && (1.97; 2.46)   && 2.66     && 2.84           &&  This work\\
\textbf{Zigzag-VLD}                          &&                &&          &&                &&        \\
$ZZV_{\texttt{Te}_{1}}$               && (2.95; 3.26)   && 2.76     && 2.84           &&  This work\\
$ZZV_{\texttt{Te}_{2}}$               && (3.02; 3.32)   && 2.78     && 2.86           &&  This work\\ 
$ZZV_{\texttt{W}}$                    && (2.72; 2.93)   && 2.71     && 2.88           &&  This work\\
\textbf{Single vacancy defect (SVD)}         &&                &&          &&                &&          \\
$V^{\texttt{SVD}}_{\texttt{Te}_{1}}$&& (2.31; 2.42)   && 2.72     && 2.83           &&  This work\\
                                    && (2.37; 2.53)   &&          &&                &&  Ref.\cite{Ozdemir2022}\\
																		&&     2.21       &&          &&                &&  Ref.\cite{Muechler} \\
$V^{\texttt{SVD}}_{\texttt{Te}_{2}}$&& (2.52; 2.72)   && 2.76     && 2.87           &&  This work                \\
                                    && (2.53; 2.70)   &&          &&                &&  Ref.\cite{Ozdemir2022}\\ 
$V^{\texttt{SVD}}_{\texttt{W}}$     && (1.59; 2.35)   && 2.67     && 2.85           &&  This work\\                                  
\hline\hline 
\end{tabular}
\label{table:Table 1} 
\end{table}

Next, we consider the effect of the VLD on the structural symmetry and electronic properties of the $1T'$-WTe$_{2}$ ML. We firstly examine the optimized structure of the VLD systems by evaluating the W-Te$_{1}$ and W-Te$_{2}$ bond lengths ($d_{\textbf{W-Te}_{1}}$, $d_{\textbf{W-Te}_{2}}$) near the VLD site, as presented in Table 1. Generally, due to the atomic relaxation during the structural optimization, the position of the atoms near the VLD site is sensitively displaced from that of the pristine system \cite{Absor2020, Li2019}. In the case of the armchair-VLD systems, the formation of the $ACV_{\texttt{Te}_{1}}$ breaks the inversion $I$ and two-folds screw rotation $\bar{C}_{2x}$ symmetries [Fig. 1(c)]. Here, removing a Te atom at the Te$_{1}$ site leads to the relaxation of the three W atoms surrounding the VLD site, resulting in the $d_{\textbf{W-Te}_{1}}$ (2.73 \AA) is smaller than that for the pristine system ($d_{\textbf{W-Te}_{1}}=2.74$ \AA). However, the $d_{\textbf{W-Te}_{2}}$ (2.85 \AA) is larger than that of the pristine one ($d_{\textbf{W-Te}_{1}}=2.83$ \AA). As a result, only a mirror symmetry plane ($M_{yz}$) along the extended vacancy line remains [Fig. 1(c)], leading to the fact that the symmetry of the $ACV_{\texttt{Te}_{1}}$ system belongs to $C_{s}$ point group. This point group symmetry also retains for other armchair-VLD systems ($ACV_{\texttt{Te}_{2}}$ and $ACV_{\texttt{W}}$) as depicted on Figs. 1(d)-1(e). However, the different atomic missing in both the $ACV_{\texttt{Te}_{2}}$ and $ACV_{\texttt{W}}$ leads to distinct effects on the W-Te bond length near the VLD site. For example, in the case of the $ACV_{\texttt{Te}_{2}}$ system, the absence of the Te atom at the Te$_{2}$ site results in both $d_{\textbf{W-Te}_{1}}$ (2.76 \AA) and $d_{\textbf{W-Te}_{2}}$ (2.91 \AA) being considerably larger than those observed in the $ACV{\texttt{Te}_{1}}$ and pristine systems. Conversely, the missing of a W atom in the formation of $ACV_{\texttt{W}}$ significantly reduces $d_{\textbf{W-Te}_{1}}$ (2.66 \AA), while slightly increasing $d_{\textbf{W-Te}_{2}}$ (2.84 \AA).

\begin{figure*} 
	\centering
		\includegraphics[width=1.0\textwidth]{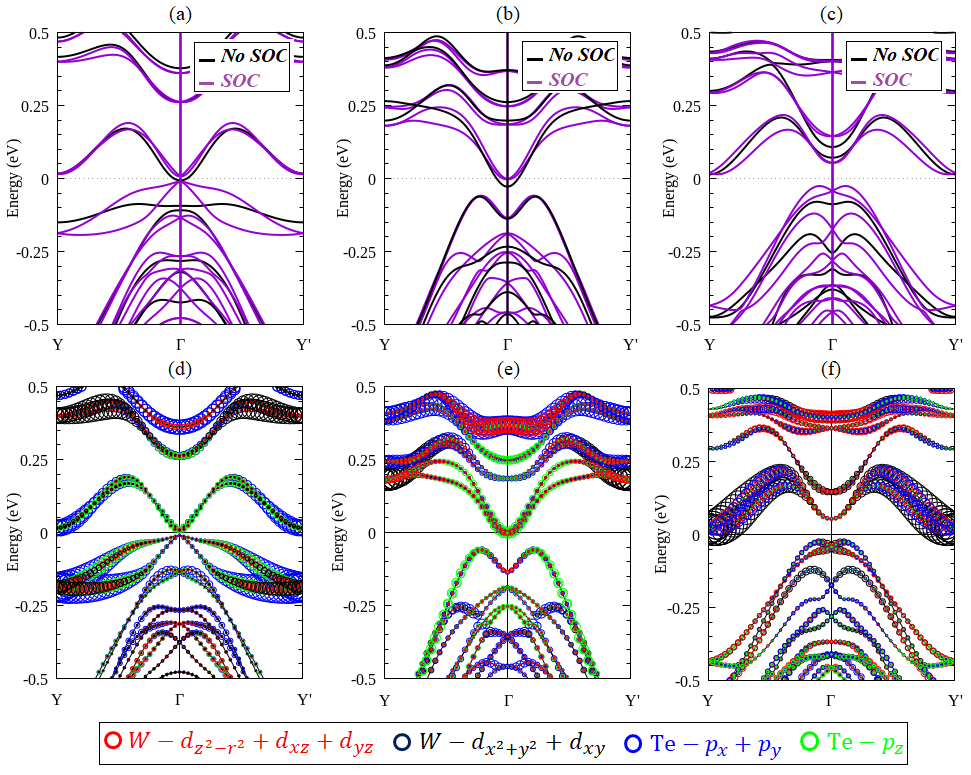}
	\caption{Electronic structures of the armchair-VLD systems. Electronic band structures calculated without (black lines) and with (red lines) the SOC for: (a) $ACV_{\texttt{Te}_{1}}$, (b) $ACV_{\texttt{Te}_{2}}$, and (c) $ACV_{\texttt{W}}$ systems are shown. Orbital-resolved projected bands calculated for: (d) $ACV_{\texttt{Te}_{1}}$, (e) $ACV_{\texttt{Te}_{2}}$, and (f) $ACV_{\texttt{W}}$ (g-i) systems are presented by the color circle, where the radii of the circles reflect the magnitude of the spectral weight of the particular orbitals to the bands.}
	\label{figure:Figure3}
\end{figure*}

Similarly, the optimized structure of the zigzag-VLD systems also undergoes significant atomic relaxation and distortion in comparison to the pristine system as shown in Figs. 1(f)-1(h). The absence of a Te atom at the Te$_{2}$ site leads to the formation of $ZZV_{\texttt{Te}_{2}}$ [Fig. 1(g)]. Here, the optimized atomic structure yields the bond lengths $d_{\textbf{W-Te}_{1}}$ and $d_{\textbf{W-Te}_{2}}$ measuring 2.78 \AA\ and 2.86 \AA, respectively, near the VLD site, which is notably larger than those observed in the pristine system. A similar atomic relaxation is also observed in the $ZZV_{\texttt{Te}_{1}}$ system [Figure 1(f)], in which the missing of the Te atom at the Te$_{1}$ site results in the substantial W-Te bond lengths [$d_{\textbf{W-Te}_{1}}=2.76$ \AA, $d_{\textbf{W-Te}_{2}}=2.84$ \AA], which are slightly smaller than those in the $ZZV_{\texttt{Te}_{2}}$ system. The smaller W-Te bond lengths ($d_{\textbf{W-Te}_{1}}$, $d_{\textbf{W-Te}_{2}}$) in the $ZZV_{\texttt{Te}_{1}}$ system compared to the $ZZV_{\texttt{Te}_{2}}$ system are consistent with those observed on the Te-based armchair VLD systems. However, resembling the structural distortion seen in the $ACV_{\texttt{W}}$ system, the formation of the $ZZV_{\texttt{W}}$ induces the relaxation of the Te atoms near the VLD sites [Fig. 1(h)], which is stabilized by a reduction in $d_{\textbf{W-Te}_{1}}$ (2.71 \AA) and an increase in $d_{\textbf{W-Te}_{2}}$ (2.88 \AA). It is noteworthy that all zigzag-VLD systems display a $C_{s}$ point group symmetry due to the presence of a single mirror symmetry plane ($M_{yz}$) perpendicular to the extended vacancy line [refer to Figs. 1(f)-1(h)].    

The structural modification induced by the presence of the VLDs significantly influenced the energetic stability of the $1T'$-WTe$_{2}$ ML, which can be confirmed by computing the formation energy $E_{f}$. As shown in Table 1, we find that both the $ACV_{\texttt{Te}{1}}$ and $ACV_{\texttt{W}}$ systems have the lowest $E_{f}$ in both the Te-rich and W-rich formations, indicating that these systems are the most favorable VLDs to be formed in $1T'$-WTe$_{2}$ ML. In addition, for both the armchair and zigzag VLD systems, we observed that the $E_{f}$ of the W-based VLD systems is lower than that of the Te-based VLD systems [see Table 1], which is in contrast to the widely studied defective systems observed on the $1H$ and $1T$ phase of the TMDCs ML. For instance, the single chalcogen vacancies and their alignment into extended line defect have been reported to exhibit the lowest $E_{f}$ as recently predicted on $1H$-MoS$_{2}$ ML \cite{Komsa, Noh, Li2019} and $1H$-WS$_{2}$ \cite{Li, Li2019}, and $1T$-PtSe$_{2}$ \cite{Absor2017, WZhang, Absor2020, Li2019}. To further analyze the stability of the VLD systems, we also compare the calculated $E_{f}$ of the VLD with those of the single vacancy defect (SVD) systems such as Te$_{1}$ SVD ($V^{\texttt{SVD}}_{\texttt{Te}_{1}}$), Te$_{2}$ SVD ($V^{\texttt{SVD}}_{\texttt{Te}_{2}}$), and W SVD ($V^{\texttt{SVD}}_{\texttt{W}}$), as presented in Table 1. Consistent with the VLD systems, we revealed that the $E_{f}$ of the $V^{\texttt{SVD}}_{\texttt{W}}$ is lower than that of the Te-based SVDs ($V^{\texttt{SVD}}_{\texttt{Te}_{1}}$, $V^{\texttt{SVD}}_{\texttt{Te}_{2}}$). The lower $E_{f}$ of the $V^{\texttt{SVD}}_{\texttt{W}}$ system compared with that of the Te-based SVD systems is also in agreement with previous results of the SVD systems observed in the bulk $1T'$ WTe$_{2}$ as reported by Ref. \cite{Song2018}.

\begin{figure*} 
	\centering
		\includegraphics[width=1.0\textwidth]{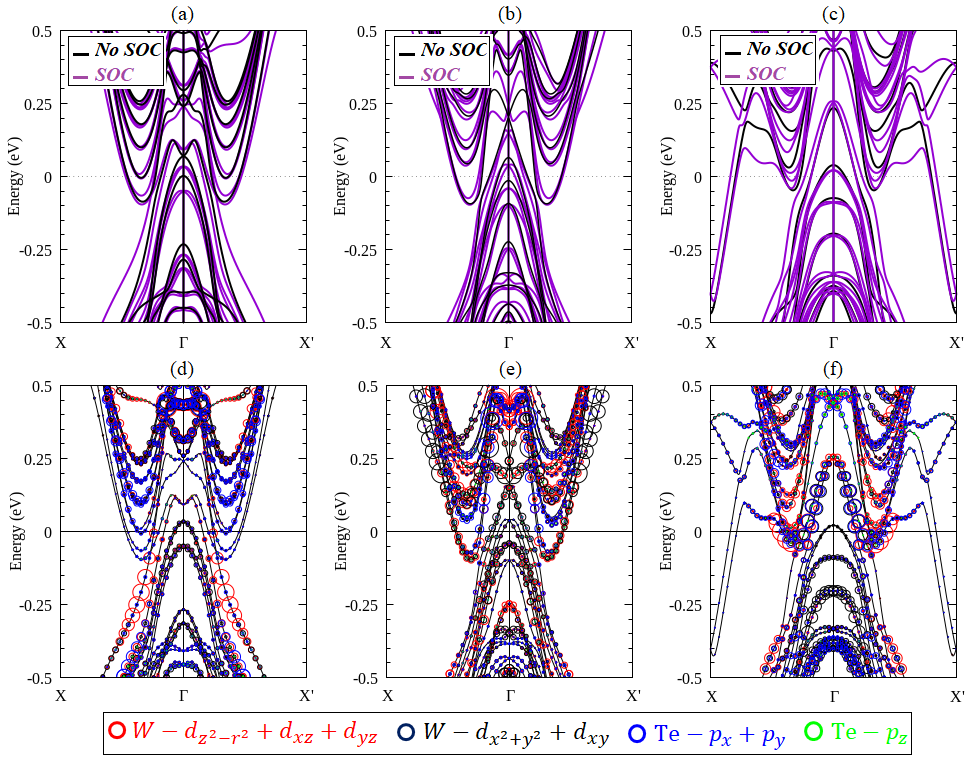}
	\caption{Electronic structures of the zigzag-VLD systems. Electronic band structures calculated without (black lines) and with (red lines) the SOC for: (a) $ZZV_{\texttt{Te}_{1}}$, (b) $ZZV_{\texttt{Te}_{2}}$, and (c) $ZZV_{\texttt{W}}$ systems are shown. Orbital-resolved projected bands calculated for: (d) $ZZV_{\texttt{Te}_{1}}$, (e) $ZZV_{\texttt{Te}_{2}}$, and (f) $ZZV_{\texttt{W}}$ (g-i) systems are presented by the color circle, where the radii of the circles reflect the magnitude of the spectral weight of the particular orbitals to the bands.}
	\label{figure:Figure4}
\end{figure*}

Figs. 3(a)-3(c) shows the electronic band structures of armchair-VLD systems calculated without and with the SOC, showcasing orbital-resolved bands projected onto atoms near the VLD site [Figs. 3(d)-3(f)]. Without the SOC, both the $ACV_{\texttt{Te}_{1}}$ and $ACV_{\texttt{Te}_{2}}$ systems exhibit metallic behavior, as evident from the crossing of the LUS to the Fermi level near the $\Gamma$ point [Figs. 3(a) and 3(b)]. However, in both systems, the HOS lies notably below the Fermi level along the $\Gamma-Y$ line, resulting in an energy gap between the LUS and HOS [Figs. 3(a) and 3(b)]. This gap is measured to be 0.1 eV and 0.04 eV for the $ACV_{\texttt{Te}_{1}}$ and $ACV_{\texttt{Te}_{2}}$ systems, respectively. Upon considering the SOC, elevated energy levels of both the LUS and HOS around the $\Gamma$ point occur owing to strong hybridization between the W-$d_{z^{2}-r^{2}}+d_{xz}+d_{yz}$ and Te-$p_{z}$ orbitals [see Figs. 3(d)-3(e)]. Consequently, both the $ACV_{\texttt{Te}_{1}}$ and $ACV_{\texttt{Te}_{2}}$ systems undergo a transition into semiconductors with small band gaps of 0.025 eV and 0.07 eV, respectively. This shift from metallic to semiconductor states, influenced by the hybridization of the out-of-plane $p-d$ orbitals, aligns with previous reports on Te-based SVD systems on the $1T'$ WTe$_{2}$ ML \cite{Ozdemir2022, Muechler}. On the other hand, the $ACV_{\texttt{W}}$ system behaves as a semiconductor with a substantial indirect band gap of 0.11 eV, where the LUS and HOS reside at the $Y$ and $\Gamma$ points, respectively [see Fig. 3(c)]. This band gap reduces to 0.075 eV when considering the SOC, attributed to the coupling between the W-$d_{x^{2}+y^{2}}+d_{xy}$ and Te-$p_{x}+p_{y}$ orbitals in the LUS around the $X$ point and coupling between the W-$d_{z^{2}-r^{2}}+d_{xz}+d_{yz}$ and Te-$p_{z}$ orbitals in the HOS around the $\Gamma$ point [see Fig. 3(f)]. In contrast to the armchair-VLD systems, all the zigzag-VLD systems exhibit a metallic state [Figs. 4(a)-4(c)]. Our orbital projection analysis indeed confirmed that the electronic states near the Fermi level are dominantly characterized by the mixing of the W-$d_{z^{2}-r^{2}}+d_{xz}+d_{yz}$, W-$d_{x^{2}+y^{2}}+d_{xy}$, and Te-$p_{x}+p_{y}$ orbitals [Figs. 4(d)-4(f)]. 

While strong coupling of the $p-d$ orbitals characterized the electronic states of the VLD systems, it also plays a significant role in determining the band splitting particularly near the Fermi level [Figs. 3(a)-3(c); Figs. 4(a)-4(c)]. It is generally understood that the coupling between atomic orbitals contributes to the nonzero SOC matrix element through the relation $\xi_{l}\left\langle \vec{L}\cdot\vec{S}\right\rangle_{u,v}$, where $\xi_{l}$ is angular momentum resolved atomic SOC strength with $l=(s,p,d)$, $\vec{L}$ and $\vec{S}$ are the orbital angular momentum and Pauli spin operators, respectively, and $(u,v)$ is the atomic orbitals. Consequently, only the orbitals with a nonzero magnetic quantum number ($m_{l}\neq 0$) will contribute to the spin splitting. For example, in the case of the $ACV_{\texttt{Te}_{1}}$ system, the hybridization between $d_{x^{2}+y^{2}}+d_{xy}$ ($m_{l}\pm 2$) orbitals of W atoms and $p_{x}+p_{y}$ ($m_{l}\pm 1$) orbitals of the Te atoms around the VLD sites induces large spin-splitting in the HOS bands at the $\vec{k}$ point along the $\Gamma-Y$ line [Fig. 3(e)]. Similar hybridization patterns inducing large spin splitting are also observed in other armchair-VLD systems ($ACV_{\texttt{Te}_{2}}$ and $ACV_{\texttt{W}}$) at the LUS bands near the $Y$ point [Figs. 3(d) and 3(f)] and the zigzag-VLD systems ($ZZV_{\texttt{Te}_{1}}$, $ZZV_{\texttt{Te}_{2}}$, and $ZZV_{\texttt{W}}$) at the LUS around the $\Gamma$ point [Figs. 4(d)-4(f)]. These results are consistent with the established understanding that strong coupling of the in-plane $p-d$ orbitals plays a pivotal role in inducing significant spin splitting in defect states, as previously reported in various TMDCs ML \cite{Li, Absor2017, Absor2018, Li2019, Absor2020}. 

To further analyze the spin-splitting properties of the VLD systems, we then focus on the two most stable VLD systems ($ACV_{\texttt{Te}_{1}}$ and $ACV_{\texttt{W}}$) since they have the lowest formation energy $E_{f}$. We identify the largest spin-splitting energy in the bands at the HOS ($\Delta E_{\texttt{HOS}}$) and LUS ($\Delta E_{\texttt{LUS}}$) near the Fermi level as highlighted in Figs. 5(a) and 5(b), respectively. In the case of the $ACV_{\texttt{Te}_{1}}$ system, we find that the calculated $\Delta E_{\texttt{HOS}}$ is 0.14 eV attained in the HOS, whereas $\Delta E_{\texttt{LUS}}$ diminishes to 0.04 eV in the LUS [Fig. 5(a)]. Conversely, when considering the $ACV_{\texttt{W}}$ system, the HOS displays maximum spin-splitting energy $\Delta E_{\texttt{HOS}}$ of 0.037 eV, whereas the LUS exhibits higher spin-splitting energy ($\Delta E_{\texttt{LUS}}=0.07$ eV) [Fig. 5(b)]. The stronger coupling between the W-$d_{x^{2}+y^{2}}+d_{xy}$ and Te-$p_{x}+p_{y}$ orbitals in the HOS of the $ACV_{\texttt{Te}_{1}}$ system [Fig. 3(a)] is responsible for inducing the larger $\Delta E_{\texttt{HOS}}$, while the mixing of the W-$d_{z^{2}-r^{2}}+d_{xz}+d_{yz}$ and Te-$p_{z}$orbitals in the HOS of the $ACV_{\texttt{W}}$ system [Fig. 3(c)] contributes minimally to the $\Delta E_{\texttt{HOS}}$. The same mechanism also explains the magnitude of the $\Delta E_{\texttt{LUS}}$ in both the $ACV_{\texttt{Te}_{1}}$ and $ACV_{\texttt{W}}$ systems. In addition, we also identified the obvious Rashba splitting around the $\Gamma$ point in both the HOS and LUS [Figs. 5(a)-(b)]. We then quantify the strength of the Rashba splitting by calculating the Rashba parameter ($\alpha_{R}$) through the relation, $\alpha_{R}=2E_{R}/k_{R}$, where $E_{R}$ and $k_{R}$ are the parameters representing the Rashba energy and momentum offset defined schematically in Fig. 5(c). Here, $E_{R}$ and $k_{R}$ parameters are important to stabilize spin precession and achieve a phase offset for different spin channels in the S-FET device. It is revealed that the calculated Rashba parameters in both the HOS and LUS bands are $\alpha_{R}^{\texttt{HOS}}=3.47$ eV\AA, $\alpha_{R}^{\texttt{LUS}}=0.21$ eV\AA\ for the $ACV_{\texttt{Te}_{1}}$ system and $\alpha_{R}^{\texttt{HOS}}=1.49$ eV\AA, $\alpha_{R}^{\texttt{LUS}}=0.07$ eV\AA\ for the $ACV_{\texttt{W}}$ system. We summarize all the calculated results of the spin-splitting parameters including the $\Delta E_{\texttt{HOS}}$, $\Delta E_{\texttt{LUS}}$, $\alpha_{R}^{\texttt{HOS}}$, and $\alpha_{R}^{\texttt{LUS}}$ in both the $ACV_{\texttt{Te}_{1}}$ and $ACV_{\texttt{W}}$ systems in Table II and compare these results with a few selected defective TMDCs ML systems from previously reported calculations. In particular the calculated values of $\Delta E_{\texttt{HOS}}$ and $\alpha_{R}^{\texttt{HOS}}$ at the HOS of the $ACV_{\texttt{Te}_{1}}$ system are the largest among the defective systems observed on the $1H$ and $1T$ TMDCs MLs \cite{Absor2020, Li2019, Absor2017, Absor2018, Li, Gupta}.  Remarkably, the associated SOC parameters discovered in the HOS and LUS of the VLDs are ample to enable the functionality of spintronics at room temperature.

\begin{figure*}
	\centering
		\includegraphics[width=1.0\textwidth]{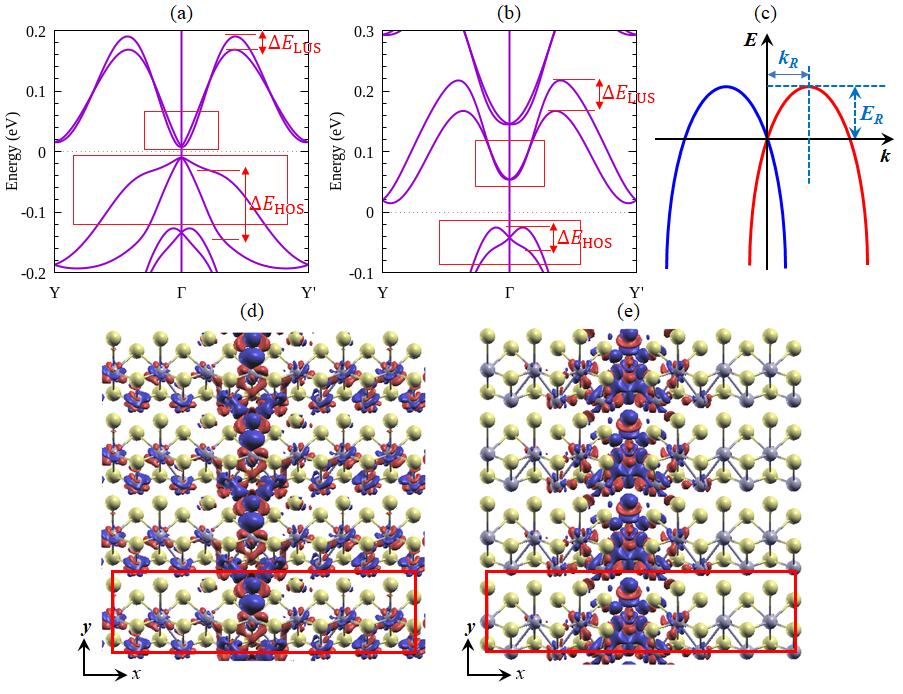}
	\caption{Spin-split bands of the defect states along the $Y-\Gamma-Y$ line around the Fermi level calculated for the VLD systems: (a) $ACV_{\texttt{Te}_{1}}$ and (b) $ACV_{\texttt{W}}$ are presented. Here, the maximum spin-splitting of $\Delta E_{\texttt{HOS}}$ and $\Delta E_{\texttt{LUS}}$ are highlighted in the HOS and LUS, respectively. (c) The schematic Rashba spin-splitting near the degenerate states around the $\Gamma$ point characterized by the Rashba energy $E_{R}$ and the momentum offset $k_{R}$. The isosurface map of the different charge density calculated for: (d) $ACV_{\texttt{Te}_{1}}$ and (e) $ACV_{\texttt{W}}$ is shown, where 0.005 was used as isovalue of the different electron density. The charge density is delocalized along the extended line defect (along the $y$-direction), forming a quasi-1D conducting channel.}
	\label{figure:Figure5}
\end{figure*}

\begin{table}[ht!]
\caption{Spin-splitting parameters including $\Delta E_{\texttt{HOS}}$ (eV), $\Delta E_{\texttt{HOS}}$ (eV), $\alpha_{R}^{\texttt{HOS}}$ (eV\AA), and $\alpha_{R}^{\texttt{LUS}}$ (eV\AA) calculated for the $ACV_{\texttt{Te}_{1}}$ and $ACV_{\texttt{W}}$ systems compared with those observed on a few selected defective TMDCs ML systems.} 
\centering 
\begin{tabular}{cc cc cc cc cc cc} 
\hline\hline 
  Systems && $\Delta E_{\texttt{HOS}}$ (eV)  && $\Delta E_{\texttt{LUS}}$ (eV) && $\alpha_{R}^{\texttt{HOS}}$ (eV\AA)    &&  $\alpha_{R}^{\texttt{LUS}}$ (eV\AA)    && Reference \\ 
\hline 
 \textbf{VLD in $1T'$-WTe$_{2}$ ML}        &&           &&            &&             &&               &&      \\
 $ACV_{\texttt{Te}_{1}}$      &&    0.14   && 0.04       &&    3.47     &&   0.21        &&  This work    \\
 $ACV_{\texttt{W}}$           &&    0.037  && 0.07       &&    1.49     &&   0.07        &&  This work    \\ 
 \textbf{VLD in $1T$/$1H$ TMDCs ML}  &&           &&            &&             &&               &&     \\
 Se-VLD $1T$-PtSe$_{2}$ 	  &&           &&            &&    1.14     &&   0.2         && 	 Ref. \cite{Absor2020} \\
 Double VLD $1H$-WSe$_{2}$  &&           &&            &&             &&	 0.14 - 0.26 &&    Ref. \cite{Li2019} \\
 Chain doped W$X_{2}$       &&           &&            &&             &&   0.0 - 0.74  &&	    Ref. \cite{Gupta}\\
 Chain doped Mo$X_{2}$      &&           &&            &&             &&	 0.1 - 1.0   &&     Ref. \cite{Gupta}\\ 
 \textbf{SVD in $1T$/$1H$ TMDCs ML}  &&           &&           &&             &&                &&   \\
 Se-SVD $1T$-PtSe$_{2}$ 	  &&    0.005  && 0.15      &&             &&	               &&   Ref. \cite{Absor2017} \\
 Se-SVD $1H$-WS$_{2}$       &&           && 0.19      &&             &&	               &&   Ref. \cite{Li}\\
 Halogen-doped $1T$-PtSe$_{2}$  &&       &&           &&             &&	   0.01 - 1.7  &&   Ref. \cite{Absor2018}\\
 
\hline\hline 
\end{tabular}
\label{table:Table 2} 
\end{table}

It is important to note here that the spin-split bands with large spin-splitting energy are achieved in the HOS and LUS near the Fermi level, which is strongly dispersive along the $\Gamma-Y$ direction. This dispersive nature of the spin-split bands is strongly different from that observed in the SVD systems. Generally, most of the SVD systems exhibit dispersionless midgap defect states which occur due to the localized charge density around the vacancy site \cite{Li, WZhang}. In contrast to our VLD systems, the strong interaction between neighboring W and Te atoms near the VLD site along the extended vacancy line leads to the dispersive spin-split bands at the HOS and LUS along the $\Gamma-Y$ direction [Figs. 5(a)-5(b)]. The strong dispersive nature of the spin-split bands at the HOS and LUS can be further clarified by the calculated results of the different charge densities presented in Figs. 5(d)-5(e) for the $ACV_{\texttt{Te}_{1}}$ and $ACV_{\texttt{W}}$ systems, respectively. We find that both the VLD systems have delocalized charge density along the extended defect line, forming a quasi-1D conducting channel. This indicates that the 1D confined defect states emerge, implying that band-like charge transport and higher mobility of the spin-split defect states are expected to be achieved \cite{Fishchuk, Gupta}, which is useful in transport-based electronic and spintronic devices.

\begin{figure*}
	\centering
		\includegraphics[width=1.0\textwidth]{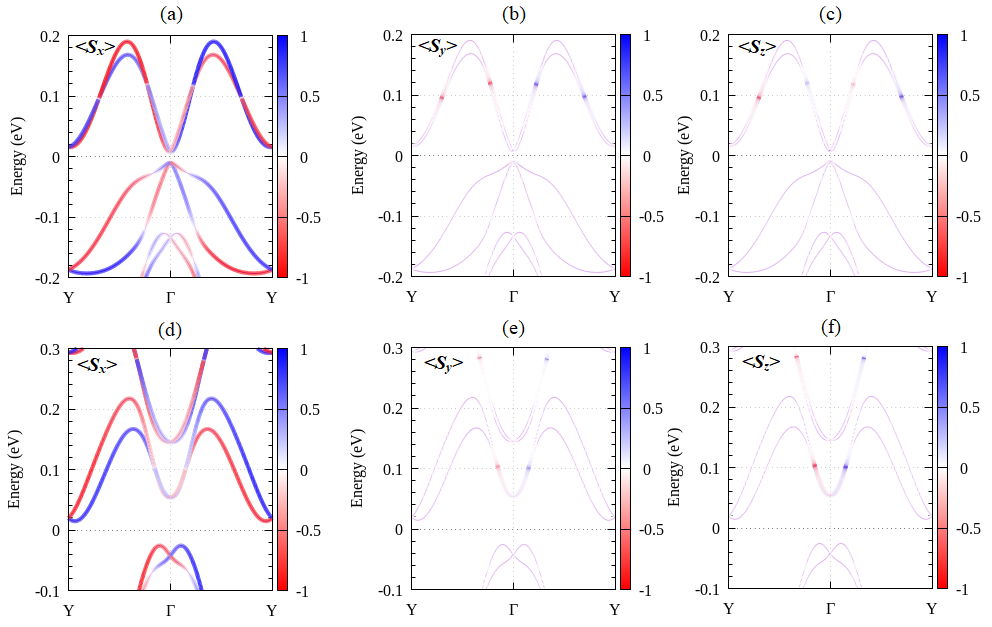}
	\caption{Spin-resolved projected bands calculated for $ACV_{\texttt{Te}_{1}}$ (a-c) and $ACV_{\texttt{W}}$ (d-f) are shown. $\left\langle S_{x}\right\rangle$, $\left\langle S_{y}\right\rangle$, and $\left\langle S_{z}\right\rangle$ represent the expected values of the spin components as indicated by the color bar.}
	\label{figure:Figure6}
\end{figure*}

To further characterize the nature of the spin-split defect states in both the LUS and HOS, we show in Figs. 6(a)-(c) and 6(d)-(f) the spin-resolved projected bands for the $ACV_{\texttt{Te}_{1}}$ and $ACV_{\texttt{W}}$ systems, respectively. It is revealed that both the VLDs exhibit the spin polarizations which are mostly characterized by $S_{x}$ component of spin polarization along the entire path of the $\Gamma-Y$ line, while the other components ($S_{y}$, $S_{z}$) are almost zero. The existence of the in-plane mirror symmetry $M_{yz}$ in both the $ACV_{\texttt{Te}_{1}}$ and $ACV_{\texttt{W}}$ systems enforces the spin components to hold the relation $\left(S_{x}, S_{y}, S_{z}\right) \rightarrow \left(S_{x}, -S_{y}, -S_{z}\right)$, which satisfy that only the in-plane $S_{x}$ component is preserved. This indicates that a perfectly collinear spin configuration in the $k$-space is achieved, which is oriented uniformly along the $x$-direction. The appearance of the collinear spin configuration enforced by the in-plane mirror symmetry results in persistent spin textures as recently observed on bulk lead-free non-magnetic materials \cite{Sheoran}, bulk WSi$_{2}$N$_{4}$ family \cite{Sheoran2023}, and various 2D materials \cite{Sasmito, Absor2019A, Guo2023, Mohanta}. This peculiar pattern of the spin configuration gives rise to the spatially periodic mode of the spin polarization, which protects the spin from decoherence and induces an extremely long spin lifetime \cite{Dyakonov, Schliemann, Bernevig, SchliemannJ, Altmann}. 

To clarify the origin of the unidirectional spin polarization observed in the $ACV_{\texttt{Te}_{1}}$ and $ACV_{\texttt{W}}$ systems, we then apply $\vec{k}\cdot\vec{p}$ perturbation theory based on the group theory analysis. This approach enables us to derive an effective Hamiltonian for the spin-split states \cite{Vajna}, as we applied recently to various defective systems in the 2D materials \cite{Absor2017, Absor2018, Absor2019, Absor2020}. As previously mentioned, the pristine $1T'$-WTe$_{2}$ ML is centrosymmetric belonging to the $C_{2h}$ point group. However, with the introduction of the VLDs, the structural symmetry is reduced. Due to time reversibility in the VLD systems, Kramer’s degeneracy is conserved at the high symmetry $k$ points in the FBZ such as $\Gamma$ and $Y$ points. Nevertheless, away from these time-reversal-invariant points, Kramer's doublet is perturbed due to the presence of SOC, which can be described through a $\vec{k}\cdot\vec{p}$ Hamiltonian. To further derive the effective SOC Hamiltonian, our initial focus centers on the vicinity of the $\Gamma$ point, before extending our analysis to encompass all paths along the $\Gamma-Y$ line, where the symmetry group aligns with that of the $\Gamma$ point. 

Using the method introduced by Vajna. $et$. $al$., the SOC Hamiltonian, denoted as $\hat{H}_{\textbf{SOC}}$, can be obtained through the following invariant expression \cite{Vajna}: 
\begin{equation}
\label{2d}
H_{\textbf{SOC}}(\vec{k})= \alpha\left(g\vec{k}\right)\cdot \left(\det(g)g\vec{\sigma} \right),
\end{equation}
where $\vec{k}$ and $\vec{\sigma}$ are the electron's wavevector and spin vector, respectively, and $\alpha\left(g\vec{k}\right)=\det(g)g\alpha\left(\vec{k}\right)$, where $g$ is the element of the point group characterizing the small group wave vector $G_{\vec{Q}}$ of the high symmetry point $\vec{Q}$ in the first Brillouin zone. By sorting out the components of $\vec{k}$ and $\vec{\sigma}$ according to irreducible representation (IR) of $G_{\vec{Q}}$, we can decompose again their direct product into IR. Notably, based on Eq. (\ref{2d}), only the IR components that are totally symmetric in this decomposition contribute to $\hat{H}_{\textbf{SOC}}$. Consequently, with the help of the corresponding tables of the point group, one can easily construct the possible term of $\hat{H}_{\textbf{SOC}}$.

In our VLD systems ($ACV_{\texttt{Te}_{1}}$, $ACV_{\texttt{W}}$), the structural symmetry is reduced to $C_{s}$ point group consisting of two elements: identity operation $E:(x,y,z)\rightarrow(x,y,z)$ and mirror symmetry operation $M_{yz}: (x,y,z)\rightarrow(-x,y,z))$. By using the character table and direct product table of the $C_{s}$ point group, the component of the wave vector $\vec{k}$ and spin vector $\vec{\sigma}$ can be classified into the IRs of the $C_{s}$ point group. By applying the table of the direct product of the $C_{s}$ point group, the possible combination of the $\vec{k}$ and $\vec{\sigma}$ components up to $n$th-order in $\vec{k}$ can be obtained, thus the effective SOC Hamiltonian $\hat{H}_{\textbf{SOC}}$ up to $n$th-order in $\vec{k}$ near the $\Gamma$ point can be written as [see supplementary materials \cite{Supporting} for the detail derivation which includes reference \cite{Vajna} therein],  
\begin{equation}
\begin{aligned}
\label{3}
H_{\textbf{SOC}}(\vec{k})= \alpha_{1}k_{y}\sigma_{x}+ \beta_{1}k_{x}\sigma_{y}+\gamma_{1}k_{x}\sigma_{z}+\delta_{1}k_{z}\sigma_{x}\\
+\alpha_{3}k^{3}_{y}\sigma_{x} +\beta_{3}k^{3}_{x}\sigma_{y}+\gamma_{3}k^{3}_{x}\sigma_{z}+ \delta_{3}k^{3}_{z}\sigma_{z}\\
+...+\alpha_{n}k^{n}_{y}\sigma_{x} + \beta_{n}k^{n}_{x}\sigma_{y} +\gamma_{n}k^{n}_{x}\sigma_{z}+ \delta_{n}k^{n}_{z}\sigma_{x},
\end{aligned}
\end{equation}
where $\alpha_{n}$, $\beta_{n}$, $\gamma_{n}$, and $\delta_{n}$ are the $n$th-order in $\vec{k}$ SOC parameters. However, due to the 1D nature of the spin-split states along the extended defect line ($y$-direction), all the terms containing $k_{x}$ and $k_{z}$ components should vanish. Therefore, Eq. (\ref{3}) can be rewritten as
\begin{equation}
\label{3a}
H_{\textbf{SOC}}(\vec{k})= \alpha_{1}k_{y}\sigma_{x}+\alpha_{3}k^{3}_{y}\sigma_{x} +\alpha_{5}k^{5}_{y}\sigma_{x}+...+\alpha_{n}k^{n}_{y}\sigma_{x}
\end{equation}

The Eq. (\ref{3a}) reveals that $\hat{H}_{\textbf{SOC}}$ is primarily defined by the $\sigma_{x}$ term, indicating a uniform spin polarization aligned along the $x$-axis near the $\Gamma$ point. This observation is consistent with the findings in our spin-resolved projected bands depicted in Fig. 6, which were obtained through DFT calculations. Furthermore, when we move away from the $\Gamma$ point to the $\vec{k}$ point along the $\Gamma-Y$ direction, higher-order $\vec{k}$ terms come into play. Nevertheless, as demonstrated in Eq. (\ref{3a}), the $\sigma_{x}$ component of the spin vector remains conserved even at higher-order $\vec{k}$ terms, preserving the unidirectional spin polarization at larger wave vectors $\vec{k}$ within the FBZ. 

\begin{table}[ht!]
\caption{Rashba parameters [$\alpha_{1}$ (eV\AA), $\alpha_{3}$ (eV\AA$^{3}$)] of the VLD systems ($ACV_{\texttt{Te}_{1}}$, $ACV_{\texttt{W}}$) obtained by fitting DFT bands of the VLDs at the HOS and LUS up to the third order term $\vec{k}$ of the Eq. (\ref{3a}). The carrier effective mass, $m^{*}$, is also obtained from the fitting of the DFT bands, which is calculated in the unit of $m_{0}$, where $m_{0}$ is the electron rest mass. The wavelength of the spin-polarized states, denoted as $L_{\texttt{SP}}$ (in nm) are also estimated by the relations, $L_{\texttt{SP}}=\left(\pi \hbar^{2}\right)/\left(m^{*}\alpha_{1}\right)$. } 
\centering 
\begin{tabular}{cccc cccc cccc cccc cccc} 
\hline\hline 
 Spin-split states &&&& $m^{*}$ ($m_{0}$)  &&&& $\alpha_{1}$ (eV\AA) &&&& $\alpha_{3}$ (eV\AA$^{3}$) &&&&  $L_{\texttt{SP}}$ (nm)   \\ 
\hline 
 $ACV_{\texttt{Te}_{1}}$ system      &&&&     &&&&             &&&&              &&&&                \\
                   HOS      &&&&  -1.1     &&&&   3.64      &&&&  -64.94       &&&&     4.42         \\
									 LUS      &&&&   1.3     &&&&   0.25     &&&&  -122.08       &&&&     159.5         \\
 $ACV_{\texttt{W}}$ system          &&&&      &&&&             &&&&              &&&&                  \\
                   HOS      &&&&  -1.25     &&&&    1.56     &&&&   -17.31       &&&&     8.47           \\
									 LUS      &&&&   1.3     &&&&    0.09    &&&&  -37.07       &&&&     197.9          \\  
\hline\hline 
\end{tabular}
\label{table:Table 3} 
\end{table}

\begin{figure*}
	\centering
		\includegraphics[width=1.0\textwidth]{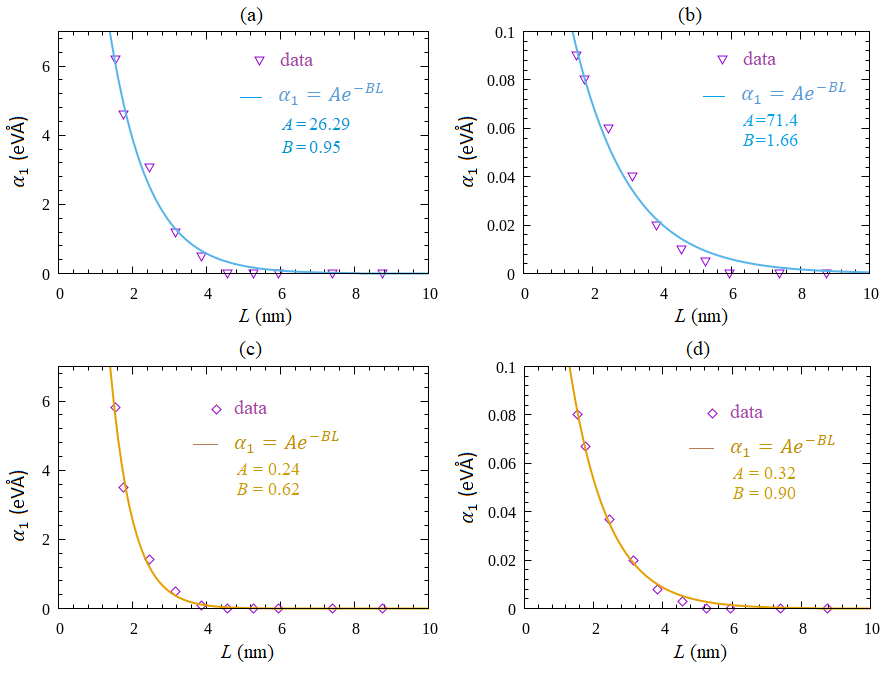}
	\caption{The variation of the $\alpha_{1}$ with the supercell length calculated at (a) HOS and (b) LUS of the $ACV_{\texttt{Te}_{1}}$ system is shown. (c)-(d) Same with (a-b) but for the $ACV_{\texttt{W}}$ system. A parametrization of $\alpha_{1}$ as a function of the supercell size $L$ along the armchair direction is performed by using the relation, $\alpha_{1}(L)=A\exp^{-BL}$, where $A$ and $B$ are the parameters obtained by fitting calculation.}
	\label{figure:Figure7}
\end{figure*}

For quantitative analysis, we performed fitting calculations of our model in Eq. (\ref{3a}) to the DFT bands in both the HOS and LUS, aiming to determine the Rashba coefficients up to third-order $\vec{k}$ terms ($\alpha_{1}$, $\alpha_{3}$); see  Fig. S1 in the Supplementary Materials \cite{Supporting} for the detailed calculations. The computed parameters are presented in Table III. For the $ACV_{\texttt{Te}{1}}$ system, we found that $\alpha_{1}=3.64$ eV\AA, $\alpha_{3}=-64.94$ eV\AA$^{3}$ for the HOS, and $\alpha_{1}=0.25$ eV\AA, $\alpha_{3}=-122.08$ eV\AA$^{3}$ for the LUS. Conversely, for the $ACV_{\texttt{W}}$ system, the values were $\alpha_{1}=1.56$ eV\AA, $\alpha_{3}=-17.31$ eV\AA$^{3}$ for the HOS, and $\alpha_{1}=0.09$ eV\AA, $\alpha_{3}=-37.07$ eV\AA$^{3}$ for the LUS. Notably, the calculated $\alpha_{3}$ parameters for the spin-split bands at the HOS and LUS in both the VLD systems are sufficiently large, indicating the significance of the third-order term in Eq. (\ref{3a}) for determining the spin-splitting characteristics of the bands. However, the computed first-order parameters $\alpha_{1}$ are closely aligned with the values obtained using the relation $\alpha=2E_{R}/k_{R}$ as shown in Table II, thereby confirming the consistency of our fitting calculations. 

Additionally, utilizing the calculated values of $\alpha_{1}$, we can estimate the spatial wavelength of the spin-polarized states using the relation $L_{\texttt{SP}}=\left(\pi \hbar^{2}\right)/\left(m^{*}\alpha_{1}\right)$ \cite{Bernevig}, where $m^{*}$ is the effective carrier mass. Here, $m^{*}$ is calculated by fitting the band dispersion around the HOS and LUS. Table III shows the calculated $L_{\texttt{SP}}$ at the HOS and LUS for the $ACV_{\texttt{Te}_{1}}$ and $ACV_{\texttt{W}}$ systems. It is revealed that the calculated $L_{\texttt{SP}}$ at the HOS of the $ACV_{\texttt{Te}_{1}}$ system is 4.42 nm, which is much smaller than that of the $ACV_{\texttt{W}}$ system ($L_{\texttt{SP}}=8.46$ nm). This value also ranks as the smallest among the VLDs systems observed on the $1H$-WSe$_{2}$ ML ($L_{\texttt{SP}} = 8.56 - 10.18$ nm) \cite{Li2019} and $1T$-PtSe$_{2}$ ML ($L_{\texttt{SP}} = 6.33 - 29.47$ nm) \cite{Absor2020}. Importantly, this value of $L_{\texttt{SP}}$ is three orders of magnitude smaller than that observed in semiconductor heterostructures ($L_{\texttt{SP}} = 5 - 10$ $\mu$m) \cite{Walser2012, Sch2014}, signifying the promise of the current system for miniaturization spintronics devices.

Next, we discussed the properties of the Rashba splitting observed in the VLD-engineered $1T'$-WTe$_{2}$ ML by considering the effect of the supercell size. We noted here that the observed Rashba splitting in the HOS and LUS of the VLD systems reflects the 1D spin-split confined defect states along the extended line defect [Figs. 5(a)-(b); Figs. 5(d)-(e)]. Consistent with the widely observed confined defect states reported on various 2D TMDCs \cite{Absor2020, Li2019, Gupta}, the appearance of the 1D spin-split confined defect states in the current system, which are mostly characterized by the $p-d$ coupling orbitals [Figs. 3(d) and 3(f), are accumulated significantly on the VLD site and penetrated rapidly into the bulk with the exponential decay as confirmed by the charge spreading in the different charge densities shown in Figs. 5(d)-5(e). Accordingly, for the VLD systems with smaller supercell sizes, the interaction between the periodic defect images strengthens the coupling of the in-plane $p-d$ orbitals in the spin-split defect states and hence increases the Rashba splitting. On the contrary, increasing the supercell size of the VLD systems decreases the periodic defect image interactions, resulting in the spin-split defect states becoming more localized with much smaller Rashba splitting. The rapid decreases of the Rashba splitting under the increasing of the supercell sizes are further well-described through the exponential dependent model. In Figs. 7(a)-7(b), we present the exponential fitting of the linear Rashba SOC parameter $\alpha_{1}$ as a function of the supercell size ($L$) calculated at the HOS and LUS for the $ACV_{\texttt{Te}_{1}}$ system, respectively, while those calculated for the $ACV_{\texttt{W}}$ system is shown in Figs. 7(c)-7(d), respectively. Here, a parametrization of $\alpha_{1}$ as a function of $L$ was performed by the following exponential function, $\alpha_{1}(L)=A\exp^{-BL}$, where $A$ and $B$ are the parameters obtained by fitting calculation; see the insert of Figs. 7(a)-7(d). As expected, we find that the calculated $\alpha_{1}$ at both the HOS and LUS tends to reduce as the thickness of the supercell of the VLD systems increases. Therefore, the Rashba splitting at the HOS and LUS can be adjusted by choosing an adequate thickness of the VLDs supercell.

Now, we discuss the possibility of realizing the VLD-engineered $1T'$-WTe$_{2}$ ML in the spintronics device applications. Here, we envision that it is possible to apply our VLD systems as an ideal 1D-Rashba channel in the SFET devices \cite{Chuang2015}. Considering the fact that the presence of the VLD systems exhibits perfectly collinear spin polarization, the spin-polarized currents can be effectively generated without any dissipation, thus increasing the performance of the SFET devices. This can be achieved by employing the lithographic patterning technique \cite{Liang2022, Dago2018}, which is compatible with current technology, to create arrays of line defects. In fact, fabrication of the 1D edge step of $1T'$-WTe${2}$ has been recently reported experimentally \cite{Peng2017, Lau}, suggesting that the realization of our VLDs systems for the device's application is now a feasible prospect.

Before concluding, we would like to discuss the potential physical implications of the observed 1D confined defect states in both the $ACV_{\texttt{Te}_{1}}$ and $ACV_{\texttt{W}}$ systems, correlated to their topological characteristics. Previously, the pristine $1T'$-WTe${2}$ ML has been reported as a quantum spin Hall insulator (QSHI) \cite{Tang2017}, featuring a quantized Hall conductance in the absence of a magnetic field originated from topologically protected 1D degenerated metallic edge states and an insulating bulk gap driven by band inversion and the strong SOC. Different from the spin-degenerate parabolic dispersion of the conventional 1D metallic systems, the QSHI host linearly dispersing 1D edge states displaying a degenerate helical-fermion mode where the spin polarity is locked to the crystal momentum (helicity) \cite{KaneMale}. In addition, the presence of the 1D helical-fermion edge states also induces the so-called Tomonaga-Luttinger liquid (TLL) behavior \cite{Jia2022}, in which the strong SOC causes the spin and chirality indices to coincide, resulting in mixed bosonic excitations involving these two degrees of freedom. In contrast to the pristine system, both the $ACV_{\texttt{Te}_{1}}$ and $ACV_{\texttt{W}}$ systems exhibit 1D spin-split polarized confined defect states where the band gap is consistently present [Figs. 4(a)-4(c)], indicating that the QSHI state does not exist. The absence of the QHSI in the spin-split confined defect states can be further confirmed by inspecting the topological nature of the electronic bands in the 2D defective bulk structure of the $ACV_{\texttt{Te}_{1}}$ and $ACV_{\texttt{W}}$ systems composed of a sequence of single square cells with point defect as shown in Fig. S2 in the Supplementary Materials \cite{Supporting}. Here, we calculate $Z_{2}$ topological invariant at the parity of the occupied bands at the time-reversal invariant points; see the Supplementary Materials \cite{Supporting} for the detailed methods, which includes references \cite{Fukui, Wanxiang, Sawahata_2019, Fu_2027} therein. Compared with the pristine system ($Z_{2}=1$), our calculation revealed that $Z_{2}=0$ for the 2D defective bulk structures, confirming that these systems are indeed trivial insulators. Even though our VLD systems did not feature the QSHI, the occurrence of the large spin-splitting bands with perfectly unidirectional spin configuration in the 1D confined defect states [Fig. 6] provides more advantages for spintronics application due to the protection of the spin from decoherence and the extremely long spin lifetime \cite{Dyakonov, Schliemann, Bernevig, SchliemannJ, Altmann}.     

\section{CONCLUSION} 

The effect of the line defect on the electronic properties of the $1T'$ WTe$_{2}$ ML has been systematically investigated by employing the first-principles DFT calculations supplemented by $\vec{k}\cdot\vec{p}$ based-symmetry analysis. We have considered six different configurations of the VLD extended  The armchair-VLDs include a Te$_{1}$ armchair-VLD ($ACV_{\texttt{Te}_{1}}$), Te$_{2}$ armchair-VLD ($ACV_{\texttt{Te}_{2}}$), and W armchair-VLD ($ACV_{\texttt{W}}$), while the zigzag-VLDs comprise a Te$_{1}$ zigzag-VLD ($ZZV_{\texttt{Te}_{1}}$), Te$_{2}$ zigzag-VLD ($ZZV_{\texttt{Te}_{2}}$), and W zigzag-VLD ($ZZV_{\texttt{W}}$), where Te$_{1}$ and Te$_{2}$ are two nonequivalent Te atoms located at the lower and higher sites in the top layer, respectively. We revealed that both the $ACV_{\texttt{Te}_{1}}$ and $ACV_{\texttt{W}}$ systems have the lowest formation energy $E_{f}$. By evaluating the two stablest VLD systems ($ACV_{\texttt{Te}_{1}}$, $ACV_{\texttt{W}}$), we have observed large spin-splitting energy in the bands near the Fermi level, which is mainly originated from the strong $p-d$ coupling of the electronic states. Importantly, we observed strong Rashba states with perfectly collinear spin configurations in the momentum space. This special spin configuration may induce a specific spin mode that protects the spin from decoherence and leads to an extremely long spin lifetime \cite{Dyakonov, Schliemann, Bernevig, SchliemannJ, Altmann}. Furthermore, we have confirmed that the observed unidirectional Rashba states are enforced by the inversion symmetry breaking and the 1D nature of the VLD, as clarified by the $\vec{k}\cdot\vec{p}$ model derived from the symmetry analysis. Our discoveries open up a potential avenue for inducing significant spin splitting in the $1T'$-WTe$_{2}$ ML, a development that holds great significance in the creation of exceptionally effective spintronic devices.

The findings presented in this article offer insights into the spin-orbit coupling (SOC) behavior within the $1T'$-WTe$_{2}$ ML, which contains the most commonly observed VLDs. These outcomes are likely applicable to other ML systems with similar structures and electronic characteristics, such as MoTe$_{2}$ ML \cite{Singh2020, Paul2020}. Specifically, the potential to control these SOC-induced spin-split states is a subject worthy of further investigation, given their high promise in spintronics applications. It would be intriguing to explore the possibility of spin reversal under the influence of an applied electric field. However, delving deeper into this topic goes beyond the scope of this paper and will be the focus of future research.

\begin{acknowledgments}

This work was supported by the Academic of Excellence Program (No. 7725/UN1.P.II/Dit-Lit/PT.01.03/2023) supported by Gadjah Mada University, Indonesia. The computation in this research was performed using the computer facilities at Gadjah Mada University, Indonesia. 

\end{acknowledgments}

\bibliography{Reference1}


\end{document}